# From hexagonal to rocksalt structure: A computational study of Gallium Selenide under hydrostatic pressure

Vo Khuong Dien[1,*]

[1]*Department of Physics, National Cheng Kung University, 701 Tainan, Taiwan*
*Email: vokhuongdien@gmail.com

**Abstract**

This article discusses the pressure-induced structural phase transition and related phonon, electronic and optical properties of hexagonal ε-GaSe using first-principles calculations. The study focuses on optimizing geometric and electronic band structures, analyzing the charge density distributions, atomic vibrations, and phonon spectra, and characterizing the optical properties of GaSe under hydrostatic pressure. The work also includes an analysis of the phase transformation mechanism using the solid-state Nudged Elastic Band (SS-NEB) method. This research sheds light on the physics of structural phase transitions in layered materials and offers potential for the development of pressure-manipulated electronics or optoelectronics.

# 1. Introduction

Group III-monochalcogenide (MX, M = Ga, In, X= S, Se, Te) is currently a topic of interest [1-3] due to their unique properties and their use in various applications such as in high-harmonic generations [4, 5], optoelectronic [6-8], photovoltaics [9, 10], and photocatalytic water splitting [11, 12]. Among these compounds, Gallium selenide (GaSe) is a central substance due to its particularly intriguing properties. The adjacent layers of GaSe are held together weakly through Van der Waals

(VdWs) interactions, whereas the bonding within each layer between the post-transition metal and chalcogenide atoms is primarily governed by covalent interactions. The interlayer interactions in GaSe have been extensively studied [13-15], and it has been found that they are relatively weak, with the cleavage energy of bulk GaSe being much smaller than that of graphite and $MoS_2$, making it easy to isolate monolayers [16-18]. According to earlier theoretical and experimental investigations, the bulk GaSe is a large electronic band gap material and undergoes a direct-to-indirect band gap transition when the number of layers decreases below the critical value [19, 20], which is an inverse tendency compared to transition metal dichalcogenides [21]. In addition to the electronic properties, layered GaSe may exhibit unique and extraordinary excitonic effects. GaSe has demonstrated strong photoresponse in recent experiments [22]. The theoretical prediction of Antonius and his co-workers indicated that the absorbance spectra of GaSe exhibit feature exciton peaks with distinct polarization selectivity, such interesting phenomena come from the symmetry of the bands in the presence of mirror symmetry [23]. Recently, theoretical calculations of Dien et al also indicated that the electronic properties and the excitonic effects of single-layer GaSe are sensitively changed with external perturbations like the mechanical strains [24].

Materials at high pressure occur at the center of the planet or star and can also be achieved in the laboratory via the diamond anvil cell (DAC) device [25]. The relative stability phases and possible transformation between them under extremely high pressure have a long-standing interest [25-29]. The knowledge about the pressure dependence on electronic and optical properties is also paramount important not only for basic sciences but also for high-tech applications [30-33]. Up to now, there are several groups have been performed on the high-pressure phase transformation of group MX-based materials utilizing ab-initio investigations [34] and high-pressure measurements [35]. Concerning GaSe, the structural phase transition from the hexagonal to the rock-salt phase has been detected to occur in a range of 21 – 25 GPa via X-ray diffraction in a DAC device [35-38]. The transition value is about ~ 20 GPa when the free-standing screw-dislocation-driven (SDD) GaSe thin film was considered [37]. The first-principles calculations of L. Ghalouci et al [34], on the other hand, found that the hexagonal to rocksalt phase transition occurred at 17.74 GPa, the electronic and optical properties of GaSe also be predicted to be very sensitive with the external hydrostatic pressure [39, 40], and thus, GaSe suitable for the development of broadband light sources and high-performance pressure sensors.

Desoite wide interest in recent years, the theoretical predictions on the pressure dependence phase transition of GaSe and their related geometric, atomic vibration, electronic and optical properties have not been undertaken. This triggers us to establish the current investigation.

In this work, the phase transition and related electronic and optical properties of the hexagonal $\varepsilon - GaSe$ will be discussed in detail. The strategy relies on the first-principles calculations with detailed analyses. The orbital hybridizations in the chemical bondings and the phase transitions will be clarified based on the optimal geometric structure, the accurate DFT and GW electronic band structures, related wave functions, and the electronic charge density distributions. The optical properties will be analyzed based on the dielectric functions, absorbance and reflectance spectra, and electron energy loss functions. In addition, the evolutions of the excitonic effects under the influence of hydrostatic pressure are also discussed in detail via the GW+BSE calculations. The phase transformation mechanism will be analyzed via the solid-state Nudged Elastic Band (SS-NEB) method.

## 2. Computational details

## 2.1 Ground state calculations

To optimize the structure and calculate electronic and optical properties, we employed the Vienna Ab-initio Simulation Package (VASP) with the Perdew-Burke-Ernzerhof (PBE) generalized gradient approximation for the exchange-correlation function in ground state calculations. We used projector-augmented wave (PAW) pseudopotentials with a cutoff energy of 500 eV to describe the electron-ion interactions, we also utilized the vdW-DFT-D2 method of Grimme et al [41], which has demonstrated reasonable accuracy for layered systems. Geometric optimization was performed using a Monkhorst-Pack sampling technique with a special k-point mesh of 25×25×4 and 15 x 15 x 15 to integrate the Brillouin zone of hexagonal and the rocksalt phases, respectively. During optimization, all atoms were allowed to fully relax until the Hellmann-Feynman force acting on each atom was less than 0.01 eV/Å, and we set the convergence condition to $10^{-8}$ eV between two consecutive simulation steps.

## 2.2 Phonon properties

Assuming that atom displacements around their equilibrium position are much smaller than the lattice constant, the problem of atomic vibrations can be addressed using harmonic approximations. The phonon dispersions and polarizations can be obtained by diagonalizing the dynamical matrix for each wave vector $\boldsymbol{k}$, given by:

$$D^{i,j}(\boldsymbol{k}) = \frac{1}{\sqrt{M_i M_j}} \sum_j C_{ij} e^{i\boldsymbol{k}\Delta R_{ij}},$$

where $C_{ij}$ represents the force constant tensor, $e^{i\boldsymbol{k}\Delta R_{ij}}$ is related to phase difference factors. The first term can be obtained from the second derivative of the total ground-state energy with respect to atomic coordinates. To obtain the force constants, equilibrium geometries, and their energy are initially determined, and finite displacements of atoms for each x, y, and z direction are obtained using phonopy codes [42]. Finally, by comparing Hellmann-Feymann forces before and after the displacement of each atom from its equilibrium position, the force constants $C_{ij}$ for all vibration modes can be achieved. Additionally, the Born effective charges (Z*) and dielectric constants ($\varepsilon$) are calculated using density functional perturbation theory (DFPT) as a correction to account for long-range electrostatic interactions in the dynamical matrix (**Figure S4**).

## 2.3 Many-Body Correlations and Optical Properties

The quasiparticle energy spectrum was obtained using the GW approximation (G0W0) [43] on the exchange-correlation self-energy, with screening effects described using the plasmon-mode model of Hybertsen and Louie [44]. The plane wave expansion was truncated at a cutoff energy of 500 eV, and response functions were truncated at 200 eV. The Brillouin zone was integrated with a 20 x 20 x 3 k-point mesh and 12 x 12 x12 k-point mesh in the Γ sampling technique for the hexagonal and the rocksalt phases, respectively. The quasi-band structure was plotted using the Wannier interpolation procedure in the WANNIER90 code [45].

To incorporate excitonic effects in optical properties, we solved the Bethe-Salpeter equation (BSE) [46] of the interacting two-particle Green's function:

$$\left(E_{c\boldsymbol{k}}^{QP} - E_{v\boldsymbol{k}}^{QP}\right) A_{vc\boldsymbol{k}}^{S} + \sum_{v'c'\boldsymbol{k}'} \langle vc\boldsymbol{k} | K^{eh} | v'c'\boldsymbol{k}' \rangle = \Omega^S A_{vc\boldsymbol{k}}^{S},$$

where, $A_{vck}^S$ is the eigenvector, $K^{eh}$ is the kernel describing the correlated electron-hole pairs, $E_{ck}^{QP}$ and $E_{vk}^{QP}$, respectively, are the excitation energies for the conduction band states and the valence band states, $\Omega^s$ is the energy of the excited state. The Tamm-Dancoff approximation (TDA) [47] was used, and energy cutoff and k-point sampling were set to resemble those in the GW calculations. A Lorentzian function with a broadening of up to 50 meV was used to replace the delta function. In this study, the excitonic effects were described using the six lowest conduction bands (CBs) and the eight highest valence bands (VBs) in the Bethe-Salpeter kernel.

# 3. Results and discussions

## 3.1 Geometric structure

Based on the stacking configuration, bulk GaSe can adopt four different phases at ambient conditions (P = 0 GPa), namely ε-(2H'/D3h), β-(2H/D6h), γ-(3R/C3v), and δ-(4H/C6v) [15], as shown in **Figure S1**. Previous experimental growths and first-principles calculations have revealed that ε-GaSe and γ-GaSe are commonly present in epitaxial films and their bulk counterpart. In this paper, we focus on investigating the structural phase transition of hexagonal ε-GaSe under hydrostatic pressure. The optimized lattice constants for ε-GaSe under vdW-DFT-D2 correction are a = 3.745 Å and c = 15.921 Å, which agree with previous experimental [48-50] and theoretical studies [34, 51]. The unit cell of ε-GaSe consists of two GaSe layers, each exhibiting a binary arsenic-like structure in the top view, while a strong covalent bond between Ga and Ga, and Ga and Se atoms forming a trigonal prismatic arrangement in the side view (**Figure 1(c)**). The Van der Waals (VdWs) interactions between the monoatomic sheets are relatively weak but still have some influence on the optimal geometric structure, as noted in **Table 1** and **Figure S2**. The high-pressure phase (rock-salt phase in **Figure 1(b)**) has a face-centered cubic structure with three orthogonal axes, where Ga and Se atoms form an octahedral arrangement (see **Figure 1(c)**).

**Table 1** Calculated and experimental lattice parameters, intralayer height $h_{intra}$, interlayer distance $h_{inter}$, the Ga-Ga bond length $d_{Ga-Se}$, the Ga-Se bond length $d_{Ga-Se}$ (in Å) and unit cell volume V$_0$ (in Å$^3$) of $\varepsilon$-GaSe at ambient conditions.

|  | a | c | $V_0$ | $h_{inter}$ | $h_{intra}$ | $d_{Ga-Se}$ | $d_{Ga-Ga}$ |
|---|---|---|---|---|---|---|---|
| LDA | 3.718[a] | 15.64[a] | 187.392[a] | 3.090[a] | 4.732[a] | 2.441[a] | 2.406[a] |
|  | 3.719[b] | 15.611[b] | 186.9[b] | 3.071[b] | 4.734[b] | 2.442[b] | 2.405[b] |
| PBE | 3.818[a] | 17.673[a] | 223.160[a] | 4.021[a] | 4.816[a] | 2.497[a] | 2.470[a] |
|  | 3.823[b] | 17.848[b] | 225.9[b] | 4.106[b] | 4.819[b] | 2.500[b] | 2.470[b] |
| PBE-D2 | 3.745[a] | 15.921[a] | 193.476[a] | 3.154[a] | 4.806[a] | 2.467[a] | 2.429[a] |
|  | 3.749[b] | 15.931[b] | 193.9[b] | 3.154[b] | 4.812[b] | 2.470[b] | 2.430[b] |
| PBE-D3 zero-damping | 3.801[a] | 16.040[a] | 200.705[a] | 3.246[a] | 4.773[a] | 2.483[a] | 2.446[a] |
| PBE-D3 Becke-Johnson | 3.774[a] | 15.876[a] | 197.256[a] | 3.173[a] | 4.801[a] | 2.445[a] | 2,477[a] |
| Xray | 3.749[c] | 15.907[c] | 193.6[c] | - | - | - | - |
|  | 3.755[d] | 15.946[d] | 194.7[d] | - | - | - | - |
|  | 3.759[e] | 15.968[e] | 195.4[e] | - | - | - | - |
|  | 3.755[f] | 15.98[f] | 195.13[f] | - | - | - | - |

a. This work
b. Ref [52]
c. Ref [53]
d. Ref [54]
e. Ref [36]
f. Ref [55]

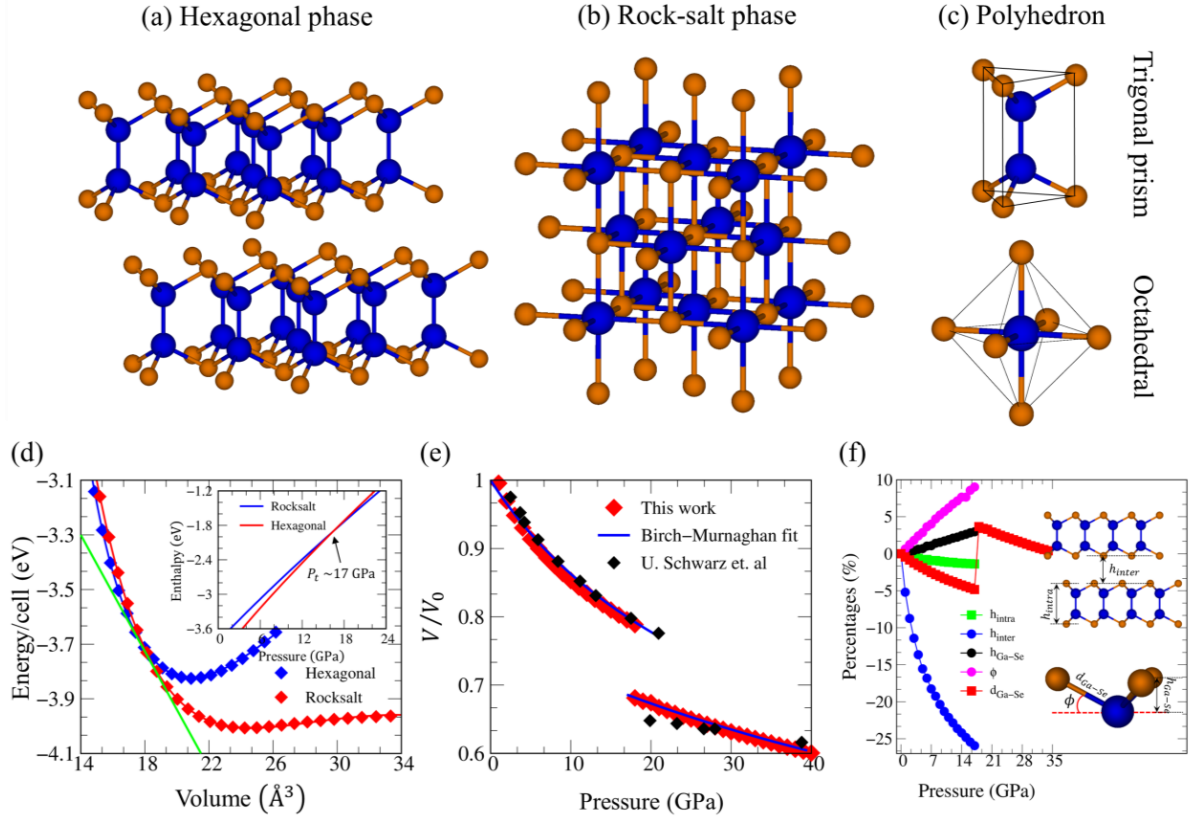

**Figure 1.** The geometric structure of GaSe with (a) hexagonal $\varepsilon$ phase (P = 0 GPa) and (b) rock-salt phase (P = 18 GPa), (c) the Ga-Se polyhedron. (d) Cohesive energy as a function of volume for the hexagonal and the rocksalt phases of GaSe, the inset Figure indicates the corresponding pressure-dependent Enthalpy. The equilibrium transition pressure $P_t$ is estimated at ~17 GPa. (e) The evolution of the lattice volume under hydrostatic pressure, the previous experiment measurement [56] is also plotted for comparisons. (f) Percent change in geometric parameters with increasing pressure, and illustrations of geometric parameters for hexagonal GaSe.

**Figure 1(d)** displays the cohesive energy curves and the enthalpy diagram of the hexagonal and rocksalt phases for $\varepsilon - GaSe$. Our results show that the transition pressure $P_t$ has a value of approximately 17 GPa, which is in good agreement with previous theoretical works [34]. The experimental works of U. Schwarz and colleagues suggested the phase transition of $\varepsilon - GaSe$ at around 21 GPa [36], the present work of Diep et al illustrated the pressure-driven hexagonal-to-rock salt transition for sub-free SDD-GaSe film at approximately ~ 20 GPa, while these for the bulk sample is in range of 22 – 25 GPa [37]. The underestimation of the transition pressure in our study compared to that of experiment observations originated from the presence of some approximations that could affect the optimized value in first-principles studies. Additionally, the computations are done for 0 K, which does not take phonon contribution into account. It is important to note that the rocksalt phase is less stable than the hexagonal phase at ambient pressure, as evidenced by its lower cohesive and enthalpy energies.

The red-square dots in **Figure 1(e)** show the changing of the unit cell's volume upon the applied hydrostatic pressure, while the solid-blue curve is the fitting results using the third-order Birch-Murnaghan Equation of State (BM EoS) [29, 57]:

$$P = \frac{3}{2}B_0\left[\left(\frac{V_0}{V}\right)^{7/3} - \left(\frac{V_0}{V}\right)^{5/3}\right] \times \left\{1 + \frac{3}{4}(B_0' - 4)\left[\left(\frac{V_0}{V}\right)^{2/3} - 1\right]\right\},$$

where $B_0, B_0', V_0$ and $V$ are, in turn, the ambient pressure bulk modulus, its pressure derivative, the fitting parameter that corresponds to the extrapolated unit cell volume of the phase being considered, and pressure-dependent cell volume. The calculated coefficients of the BM EoS for the hexagonal structure are $B_0 = 54.54 \, GPa$, and $B_0' = 2.84$, and for the rocksalt phase are $B_0 = 153 \, GPa$, and $B_0' = 3.1$. Generally, the compressing volume nearly monotonously decreases with pressure increase and gets a dramatic change at the critical pressure ~ 17 GPa. Quantitatively, the volume variation $\Delta V = |V - V_0|/V_0$ is 10% and compatible with more than 8% observed in experiments [36]. The pressure-volume diagram is in good agreement with previous experimental measurements [36] (The black square dot in **Figure 1(e)**). It is important to emphasize that the shrinking of the unit cell volume will be accompanied by an increase in the total energy of the crystal and the bulk modulus. Furthermore, the total valence bandwidth and the electronic properties are also expected sensitive to change with the applied pressure.

In the stability field of the $\varepsilon - GaSe$, the presence of intralayer height ($h_{inter}$), interlayer distance ($h_{intra}$), the Ga-Se height difference ($h_{Ga-Se}$), the Ga-Se bond length ($d_{Ga-Se}$), and the angle ($\phi$) between the Ga-Se bond and the Se atom sheet can provide insights into the deformations of GaSe under hydrostatic pressures (**Figure 1(f)**). At a pressure below 17 GPa, the interlayer distance gradually decreased up to 25% at P ~ 17 GPa mainly due to the weak vdW force, causing a gradual shrinkage of the vertical lattice constant c of GaSe [37]. However, the intralayer $h_{intra}$ remained almost constant. The GaSe chemical bond length $d_{Ga-Se}$ on the other hand, significantly reduced and got a drastic change at ~ 17 GPa, indicated the phase transition.

Despite a 5% reduction in $d_{Ga-Se}$, $h_{Ga-Se}$ increased mildly by about 3% due to an increase in the angle $\phi$ from 27.9° to 30.4°, leading to reduce the in-plane atomic interactions of Ga and Se atoms. The observed chemical modifications are expected to dramatically impact the electronic and optical properties of GaSe through alterations in orbital overlaps and separations.

## 3.2 Atomic vibrations and phonon properties

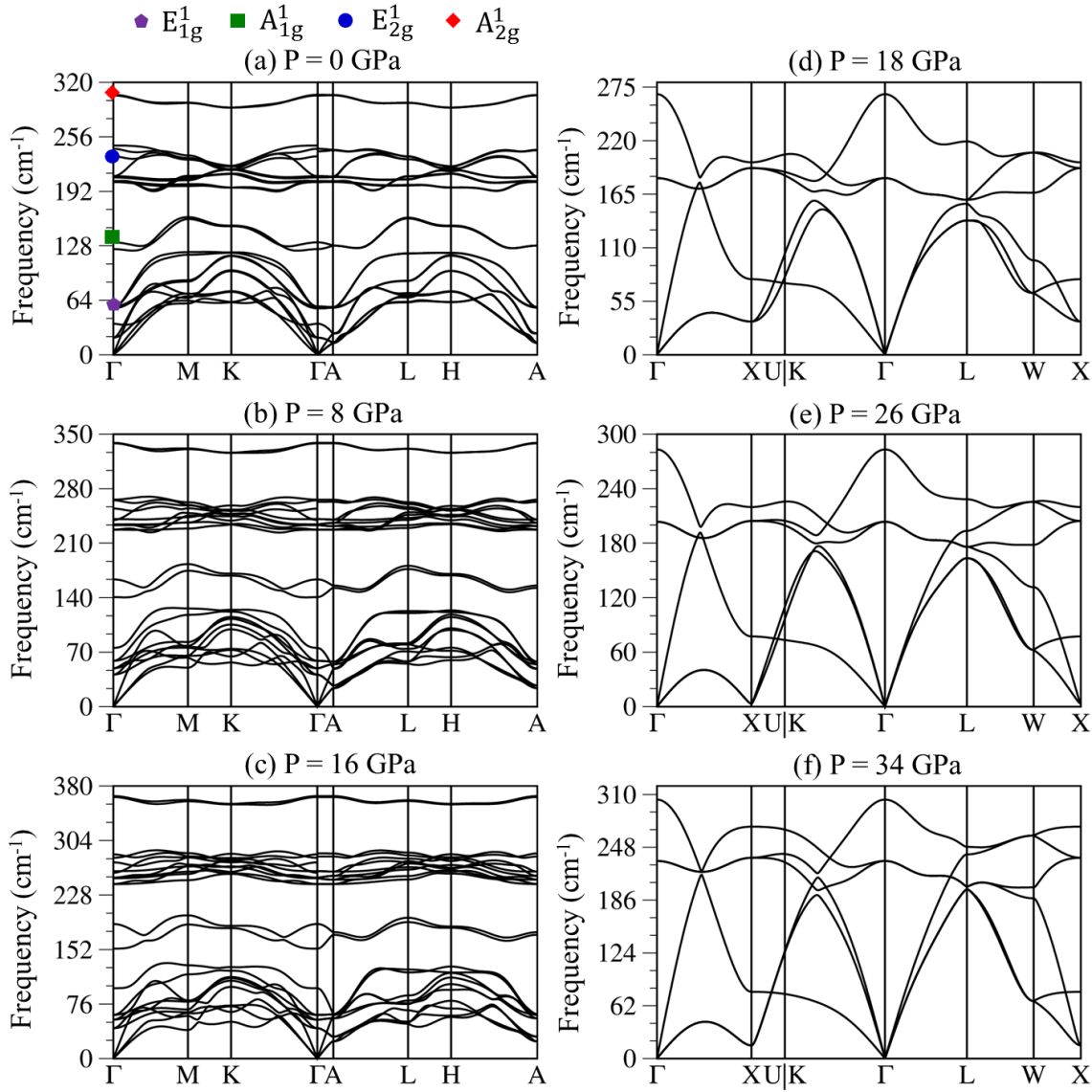

**Figure 2.** Calculated phonon dispersions for GaSe under six presentative pressures with their high-symmetry points are shown in **Figure S3**. The Raman active modes of hexagonal GaSe are marked by the color symbols.

The study of phonon energy dispersion provides valuable insights into the dynamic vibrations and structural transformation of materials. **Figure 2** displays the phonon band structure of GaSe under six representative pressures, the hexagonal ε-GaSe phase where the bulk phonon dispersion (**Figure 2a**) includes 3 acoustic and 21 optical modes, corresponding to 8 atoms in the unit cell. At long wavelength limit, two in-plane vibration acoustic modes (LA and TA) exhibit linear dispersion with higher frequency than the out-of-plane acoustic mode (ZA), the latter has a quadratic dependence similar to other layered materials such as graphite [58] and $MoS_2$ [59] and plays a significant role in the material's thermal properties at low temperatures. The VdWs interactions between the two GaSe layers give rise to low-frequency optical modes at 20.05 $cm^{-1}$ and 36.38 $cm^{-1}$, corresponding to rigid-layer shear and vertical motions, respectively. These soft vibration modes may dominate the low-temperature heat capacity [60]. It is important to note that there are large phonon band gaps between the low-frequency optical phonon branches and other optical modes, resulting from the significant difference in atomic mass between Ga and Se, as confirmed by the phonon density of states in **Figure S5**. As compression pressure increases (**Figures 2(b-c)**), the bandwidth of the GaSe phonon band structure gradually increases into higher frequencies due to the shrinkage of the vertical VdWs gap and significant shortening of the intralayer bondings. More dispersive phonon branches near the Γ point can be observed, suggesting a larger group velocity, while the LA and TA phonon branches become increasingly softer around high-symmetry points such as M, K, L, and H. These phenomena directly influence phonon accumulation and strongly modify phonon heat transport properties. **Figures 2(d-f)** show the phonon band structure of the high-pressure rocksalt phase of GaSe, which differs significantly from that discussed above. Only 6 phonon branches correspond to the 2 atoms in the rocksalt primitive cell, and the large phonon band gap observed in the hexagonal phase is entirely absent in the rocksalt structure. More dispersive phonon branches can be detected. Importantly, there is no imaginary frequency in the phonon band structure of any sample, indicating the dynamic stability of GaSe under pressure.

**Table 2** Vibration direction, calculated frequency, and the changing rate ($d\omega/dP$) of the relevant Raman active modes. The measurement data of the prior works are also listed for comparison.

| Mode parameter | Direction | Frequency ($cm^{-1}$) | $d\omega/dP$ ($cm^{-1}/GPa$) |
|---|---|---|---|
| $E_{1g}^1$ | In-plane | 56.55[a] | 0.28[a] |
| | | 58.8[b] | 0.26[b] |
| | | 60.1[c] | - |
| $A_{1g}^1$ | Out of plane | 131.85[a] | 4.01[a] |
| | | 133.1[b] | 4.96[b] |
| | | 134.6[c] | - |
| $E_{2g}^2$ | In-plane | 232.79[a] | 2.73[a] |
| | | 251.4[b] | 2.35[b] |
| $A_{1g}^2$ | Out of plane | 305.0[a] | 4.26[a] |
| | | 308.1[b] | 3.85[b] |
| | | 307.8[c] | - |

a. This work
b. Measurement data from Ref [37]
c. Measurement data from Ref [61]

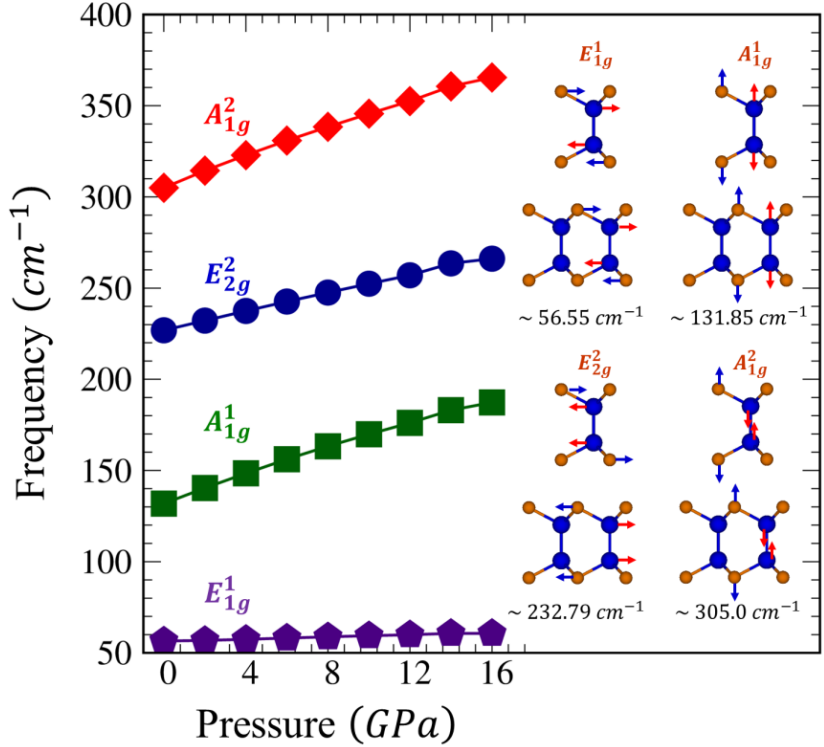

**Figure 3.** (a) The pressure-dependent Raman active modes of hexagonal $\varepsilon - GaSe$, and (b) Visualized configuration of the corresponding Raman active vibration modes.

To better understand the evolution of phonon vibrations under pressure, we plotted the active Raman modes versus pressure in **Figure 3(a)**, along with the corresponding vibration modes in **Figure 3(b)**. We also would like to emphasize that the spectra only go up to 17 GPa, as the rocksalt phase of GaSe is Raman inactive due to its inversion symmetry crystal structure [56, 62]. At the ambient pressure, the hexagonal phase of $\varepsilon - GaSe$ exhibit four distinct Raman modes: $E_{1g}^1$, $A_{1g}^1$, $E_{2g}^2$, and $A_{1g}^2$ located at 56.52 cm$^{-1}$, 131.79 cm$^{-1}$, 232.74 cm$^{-1}$, and 304.91 cm$^{-1}$, respectively. For comparison, the experimental values from the Raman measurement are also listed in **Table 1**, the overall agreement between theory and experimental is quite good, even for the out-of-plane vibration modes. As pressure increases, all of the active Raman modes increase linearly in frequency, reflecting a broadening of the phonon band structure and accompanying lattice shrinkage. The changing rates of these modes are 0.28 $cm^{-1}/GPa$, 4.01 $cm^{-1}/GPa$, 2.73 $cm^{-1}/GPa$, and $A_{1g}^2$= 4.26 $cm^{-1}/GPa$, respectively, for $E_{1g}^1$, $A_{1g}^1$, $E_{2g}^2$, and $A_{1g}^2$ Raman active modes. It is worth noting that the out-of-plane vibration modes, such as $A_{1g}^1$ and $A_{1g}^2$ modes, have a much faster changing than the others, such as $E_{1g}^1$. This is because compression, and thus the corresponding force constant is more sensitive along the c axis, which is attributed to the presence of weak vdWs forces. As the pressure increases, the distance between layers decreases, and the interlayer interaction is strongly enhanced, leading to a rapid increase in the frequency of $A_{1g}^1$ and $A_{1g}^2$ modes. In contrast, the low-frequency $E_{1g}^1$ mode, which has a frequency determined by the interlayer force associated with a Se-Se bond and the intralayer bond-bending force (**Figure 3(b)**), showed an insensitive characteristic to pressure. The little variation in frequency suggests that the large increase in the strength of the interlayer force is canceled out by the weakening intralayer Ga−Ga bond-bending force. Similar behavior was also found in the low-frequency Raman active modes of multilayer InSe [56]

## 3.3 Electronic properties

The electronic properties of $\varepsilon-$ GaSe crystals under applied pressure were examined by performing theoretical calculations of the electronic band structures. Both DFT and GW levels of theory were utilized, with the latter achieved through the Wannier90 code [63]. **Figures 4(a-f)** illustrate the electronic band structure of GaSe along the high symmetry points of the first Brillouin zone (**Figures S3**), in case of the absence of external pressure, the bulk GaSe electronic properties could be characterized by the direct energy gap of 0.73 eV at the Γ point. The gap value is consistent with previous theoretical DFT work [19, 36] but rather smaller than that of the experimental observation [64, 65] due to the absence of quasi-particles corrections. The electronic band gap of hexagonal GaSe is significantly enhanced ($E_g \sim 1.97\ eV$) since the correction of electron-electron interactions (GW approximation) has been applied (the red curve in **Figure 5(a)** and **Figure S6**), this value is in good agreement with previous measurements [65, 66]. In addition to band gap enhancement, the many-body interactions also make a significant alteration in the energy dispersion. For example, the energy difference between $CBM(\Gamma)$ and $CBM(L)$ (~10 meV) are relatively smaller than that at the DFT level of theory, and thus, the indirect or direct nature of GaSe is rather difficult to distinguish, the conclusion is in good agreement with previous works [67].

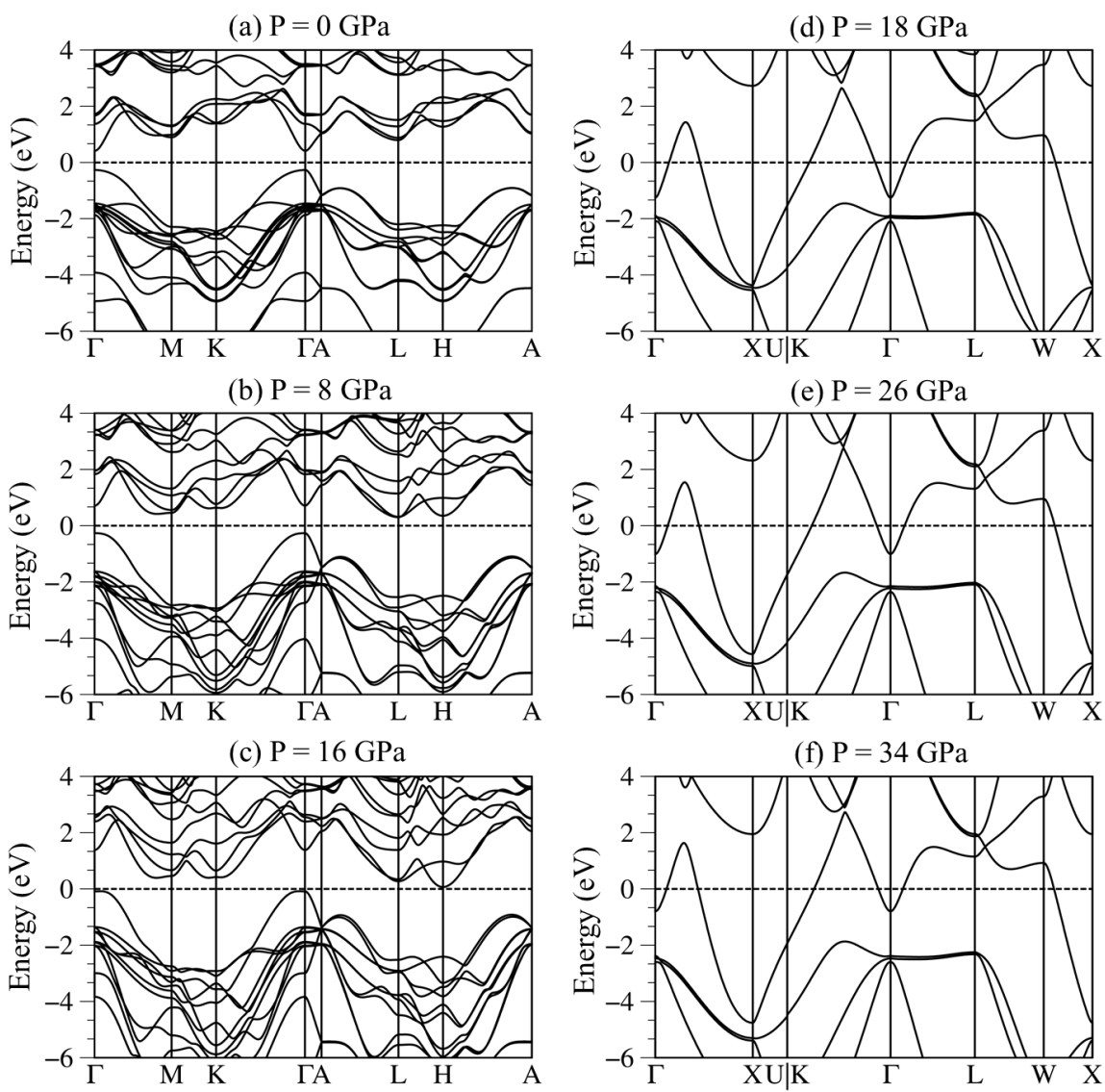

**Figure 4.** The DFT electronic band structure of GaSe along the high-symmetry points under six presentative pressure.

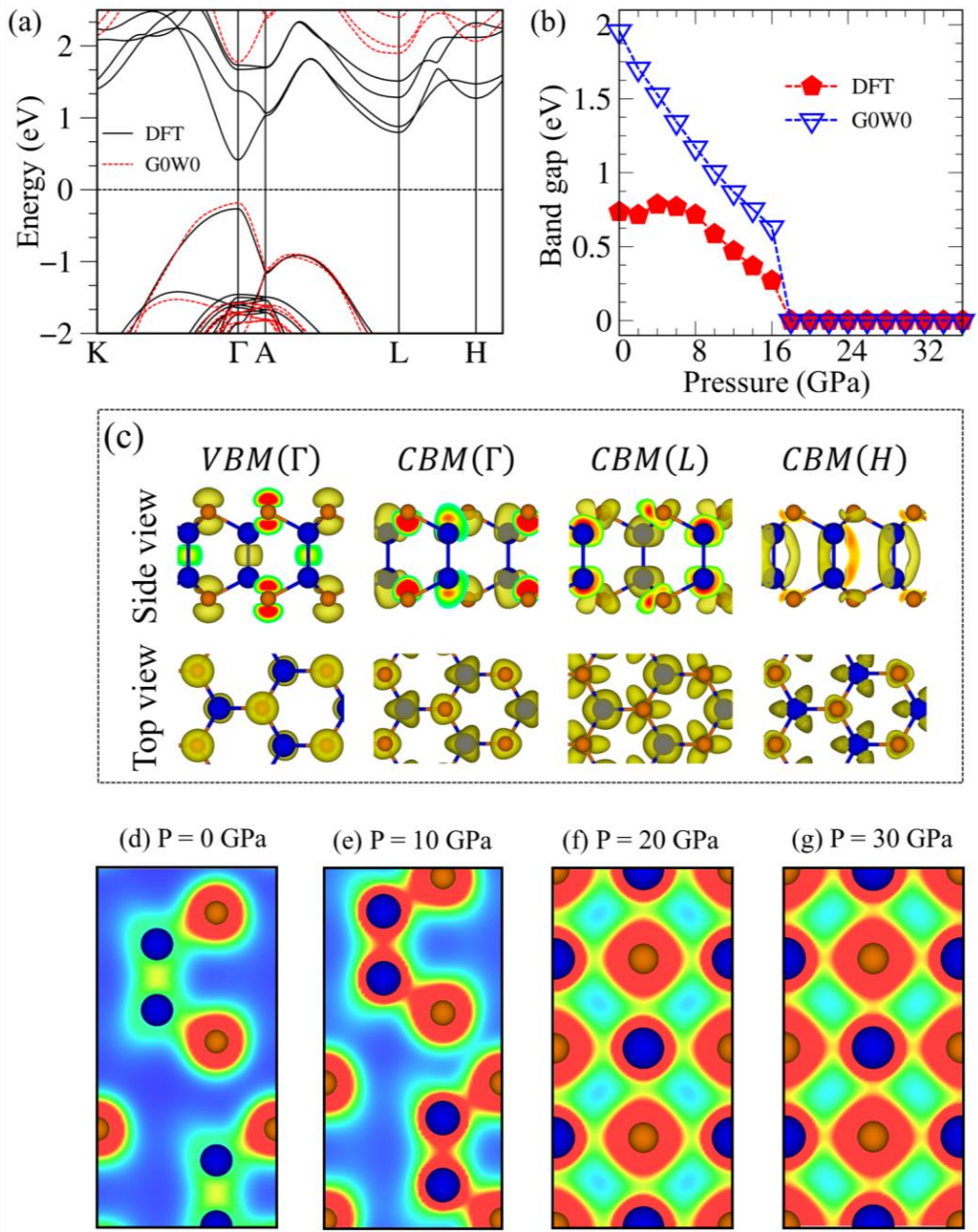

**Figure 5.** (a) A comparison of the electronic band structure using GW and DFT approximations, (b) Evolution of the band gap under DFT and GW approximations, (c) Top and side views of the band-decomposed charge densities at the critical band-edge states of the first layer of hexagonal GaSe, and (d-g) Charge density distribution of GaSe at four representative pressures.

To further comprehend pressure-induced band structure modulation, the band decomposed charge densities, and orbital-projected electronic band structures have been calculated. **Figure 5(c)** and **Figure S7** show that the VBM(Γ) is mainly composed of the out-of-plane bonding character of Ga-4$p_z$ and Se-4$p_z$ orbitals, while the CBM(Γ) is derived from the out-of-plane anti-bonding interactions of the Ga-4s and the Se-4$p_z$ charge densities. The in-plane anti-bonding character of Ga-4s and Se-(4$p_x$ + 4$p_y$) orbitals are related to the CBM(L), and the in-plane Ga-(4$p_x$+4$p_y$) and Se-(4$p_x$ + 4$p_y$) orbitals characters connected to the CBM(H). Therefore, a large modification of the band states in the vicinity of Γ related to the changing of the coupling of layers, while the interactions of in-plane orbitals of Ga and Se atoms will result in the changing of the band states at L and H.

Under pressurization, the electronic band structure may exhibit significant differences from that of the pristine, as the conduction and valence energy sub-bands become more dispersive due to the shortening of interatomic distances. This leads to greater mobility for electrons and holes. The enhancement of vertical orbital interactions due to the reduction of Se-Se interlayer/intralayer significantly increases the energy of the band edge states at VBM(Γ) and CBM(Γ), resulting in an enhanced direct band gap at the Γ point. Conversely, pressurization greatly reduces the in-plane orbital interactions since the Ga-Se vertical distance gradually increases (discussed in **Figure 1(f)**). Consequently, the new CBM is located at L, leading to a direct-indirect transition under pressure. This is consistent with previous theoretical predictions [67] and experimental measurements [36]. When the pressure is greater than 8.0 GPa, the CBM located at the H point becomes energetically more favorable compared to the CBM located at the L point. This causes a crossover of the CBM from the L point to the H point (**Figure 4(c)**). At extreme pressure, such as P ~ 17 GPa, the electronic structure of GaSe undergoes a significant change due to the phase transition. **Figure 4(d)** shows the electronic band structure of the high-pressure rock-salt phase, where there is no band gap between the occupied and unoccupied states (metallic) due to their overlapping and the continuous availability of electrons in these closely spaced orbitals. Such behavior is reflected in the spherical-extended nature of the charge density (**Figures 5(d-g)**).

The sensitivity of the electronic properties to external perturbation has been mentioned in a previous section. These features are highlighted in **Figure 5(b)**, which depicts the evolution of the bandgap of gallium selenide under pressure. It is worth noting that the bandgap adjustment is not a straightforward process but rather a complicated one. The bandgap adjustment under the DFT level of theory can be classified into four sub-progresses. The first sub-progress is that the E$_g$ increases up to a maximum electronic bandgap of 0.8 eV at a pressure of 4 GPa, owing to the direct-indirect transition. After reaching the maximum, the electronic bandgap monotonically decreases as higher strains are applied due to the wave function overlap, with a changing rate of about 0.1 eV/GPa. Under a critical strain of P = 17 GPa, the material undergoes a transition from a semiconductor to a metal as previously predicted. Beyond this pressure, the GaSe compound only exhibits metallic properties and the electronic band gap is completely absent. The band gap evolution under GW approximation follows a different trend, the electronic band gap almost monotonically decreases as the external pressure increases and the semiconductor–metal transition also occurs at 17 GPa. However, the direct-indirect transition cannot be detected on the pressure-bandgap diagram since the energy difference between $CBM(\Gamma)$ and $CBM(L)$ at ambient conditions is relatively small to make an obvious signal under the GW corrections.

The charge density distribution can provide insight into the orbital interaction and subsequent structural transformation [68, 69]. At the ambient condition (**Figure 5(d)**), the charge density is mainly located around the Se atom due to its higher electronegativity [70]. A strong deformation around both Ga and Se atoms indicates the covalent nature of the Ga-Se chemical bonding. Under external hydrostatic pressure below 17 GPa (**Figures 5(d-e)**), the charge distribution between the two atoms experiences a noticeable change. The wave functions overlap between two sub-systems is significantly increased, but the interaction between the two atoms remains covalent. Above 17 GPa (**Figure 5(f)**), the charge density distribution around Ga and Se atoms undergoes a significant transformation as a result of a new atomic arrangement, and extra Ga-Se bondings are established. Beyond this critical pressure (**Figure 5(g)**), the charge density distribution of Ga and Se atoms becomes more and more spherical-extended, which is an indication of metallization. This behavior is different from that observed in the wurtzite-to-rocksalt phase transition of MgO and GaN [27], where the charge density of their rocksalt phase is more uniformly but localized around the atoms, suggesting a particular ionic bonding.

## 3.4 Optical properties and excitonic effects

Similar to other layered materials, GaSe exhibits strong anisotropic optical properties at ambient pressure [71] due to the significant difference in lattice constant and material environment along the x-/y- and z-directions (**Figure S8**). Theoretical investigations by M. Schluter and colleagues

[72] successfully connected prominent optical excitations with atomic character at specific band edge states, while Noritaka Kuroda et al [73] investigated the contribution of phonon vibration to absorption in the infrared and far-infrared regions. Several experimental studies have also investigated the optical properties of GaSe. For example, J. Camassel [65] established the near-band-edge optical properties of GaSe at low temperatures (~10K), with low-frequency sharp peaks related to excitons. Additionally, M. Gauthier and colleagues confirmed the sensitivity of GaSe's optical properties to external pressure, and U Schwarz's recent work [36] indicated a total change in the dielectric properties of GaSe at extremely high pressure due to phase transition.

**Figure 6(a)** illustrates the imaginary part of dielectric functions $\varepsilon_2(\omega)$ of GaSe for the light polarization along the z-direction at the ambient pressure, with different levels of theory established. Within the DFT-RPA framework, the absorption spectra indicate the presence of the optical gap/threshold frequency $(E_g^o)$ at approximately 0.76 eV with another prominent peak at about 1.2 eV denoted by the red and blue triangles, respectively. These singularities arise from the vertical transition at the band edge states vicinity of the Γ point, with their transition mechanism assigned in the inset of **Figure 6(a)**. The optical spectra display a significant blue shift due to the application of GW corrections, while the opposite is true for the GW-BSE spectra as electron-hole interactions have been considered, e.g., $E_g^o \sim 1.9\ eV$. Although the interaction is weak compared to that of the single-layer system [23], the excitonic effects, evidenced by the difference in the GW-RPA and GW-BSE spectra, could be detected with the exciton binding energy $E_{xb} \approx 90\ meV$ at ambient pressure. The optical gap position and the relatively weak exciton binding of GaSe in this work agree well with previous experimental measurements [55, 65, 74]. In **Figure 6(b)**, the optical absorbance spectra of high-pressure rock-salt GaSe are presented. The absorbance spectra display a finite value at low frequency, indicating the dominance of the intraband transition of the free electrons around Γ point. Additionally, there is a prominent interband transition at approximately 2.8 eV, with the excitation mechanism shown in the inset of **Figure 6(b)**. Although electron-hole interactions affect optical spectra, their effects are almost undetectable in this case, as metallic systems with high free electron density are heavily influenced by electronic screening effects.

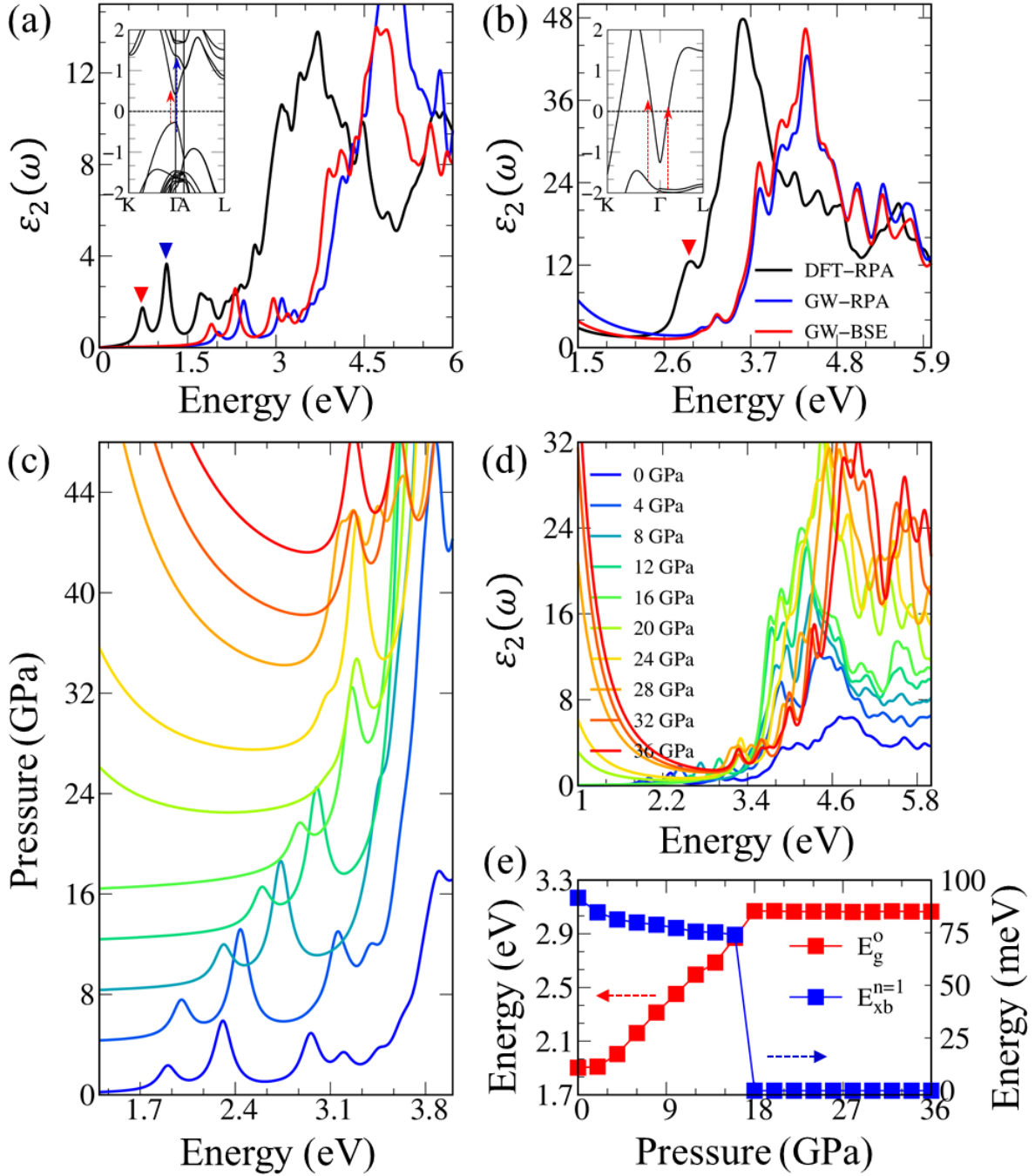

**Figure 6.** The imaginary part of the dielectric function $\varepsilon_2(\omega)$ under different levels of theory of (a) hexagonal GaSe, and (b) high-pressure rocksalt GaSe. The inset Figure indicated the transition mechanism of the first prominence interband transition. (c) The pressure-dependent optical excitation of GaSe, (d) similar plot as above but for a wider energy range ($1\ eV \leq \omega \leq 6\ eV$), and

(f) the evolution of the first interband transition and the exciton binding energy as functions of applied pressure.

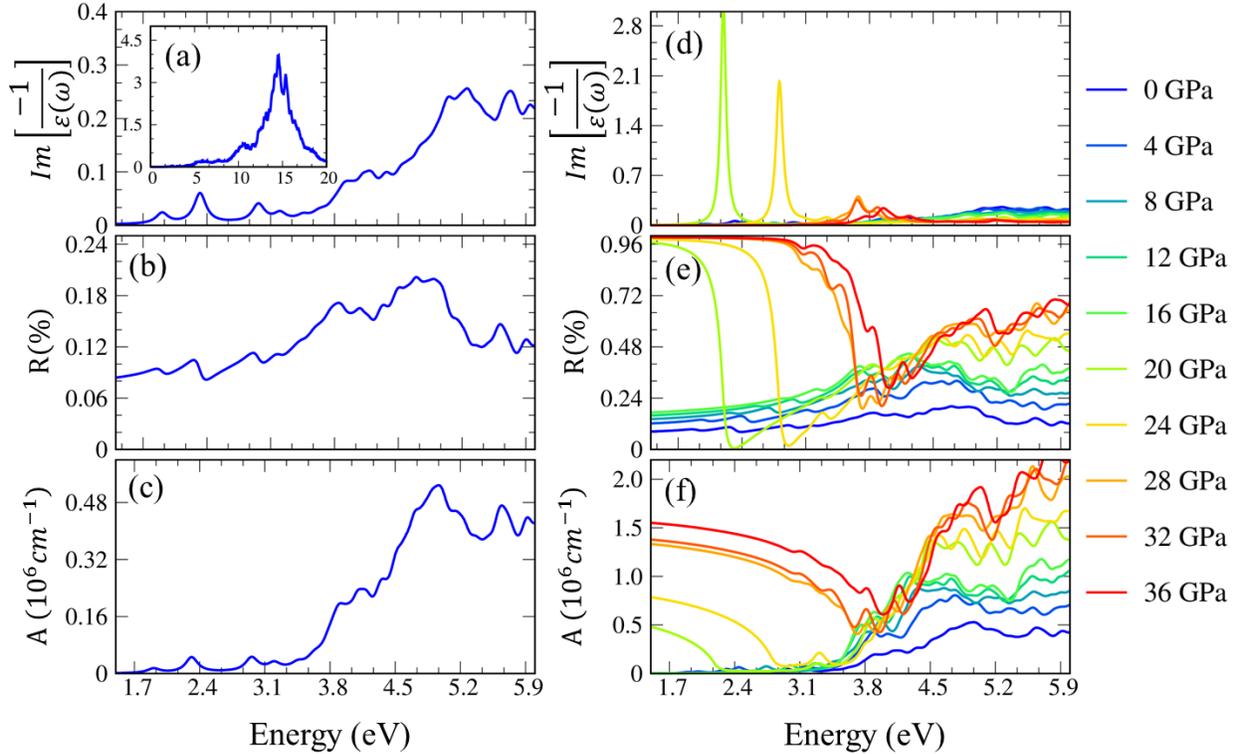

**Figure 7.** (a) The energy loss function, (b) absorbance, and (c) reflectance spectra of pristine $\varepsilon - GaSe$. (d), (e), and (f) the changing of the various optical properties under pressure.

The low-frequency imaginary part of dielectric functions $\varepsilon_2(\omega)$ is shown in **Figure 6(c)**. The absorption curves show no significant changes up to 4 GPa, but the optical spectra begin to shift to higher frequencies, reflecting an enhancement of the direct band gap at the Γ point, in agreement with the above theoretical predictions and previous experiments [39]. Beyond the critical pressure (P = 17 GPa), there is only little change in the low-frequency optical spectra. In addition to the tailoring of the low-frequency optical gap, the external pressure is also induced to enhance the optical absorbance at the high-frequency region as indicated in **Figure 6(d)**. The changing of the optical gap and the exciton binding energy is analyzed and plotted in **Figure 6(e)**, where the modulation of the exciton binding energy ($E_{xb}$) is found to be a complex process. $E_{xb}$ exhibits a remarkable change at low pressures (P < 4 GPa) due to the reduction/enhancement of the band gap/electronic screening, but then slowly decreases as higher external forces are applied. When the external pressure exceeds the critical phase transition value, the electron-hole couplings drop dramatically to ~ 0 eV, reflecting the metallic behavior of the rocksalt GaSe.

**Figures 7(a-c)** show the energy loss functions, defined as $Im[-1/\varepsilon(\omega)]$, which can provide valuable information about the charge screening ability and optical properties of materials. Previous studies [19] and insert of **Figure 7(a)** have indicated that due to the ill-defined $\pi$ and $\sigma$ valence electrons in the sp$^3$ GaSe crystal, only one strong plasmon mode at energy greater than 8 eV exists, and the low-frequency collective excitation is almost disappeared. As shown in **Figure 7(d)**, $Im[-1/\varepsilon(\omega)]$ remains almost unchanged for the pressure below 17 GPa, whereas the opposite is true for pressures beyond the critical value, such as P = 20 GPa. In such a case, the free electron contribution is denoted by a strong peak emerging below 2.4 eV, consistent with the metallic behavior of rocksalt GaSe as has been discussed. The energy and intensity of the Drude contribution undergo significant changes with higher pressure application, as evidenced by the gradual shift of strong peaks towards higher energy with simultaneous broadening/suppression of intensity.

**Figures 7(b)** and **7(c)** present the reflectance $R(\omega)$ and absorbance spectra $A(\omega)$ of GaSe. Reflectance indicates the percentage of light reflected from the front surface, while the inverse of absorbance indicates the characteristic length of penetration of an electromagnetic wave in a condensed-matter system. In the low-energy region below 3.5 eV, the reflection is weakly energy-dependent, with typical z-direction values of 10% (**Figure 7(b)**). The absorption coefficient is almost negligible due to the absence of optical excitations (**Figure 7(c)**). However, optical excitations cause significant fluctuations in reflectance and strong absorption at 4 eV, expected for optical-allowed transitions. The inverse value of the absorption coefficient mostly lies in a range of 200 Å, implying that GaSe has potential optoelectronic technology applications due to rich optical excitations, allowing easy absorption of photon beams penetrating the medium. To fully comprehend the evolution of the optical properties of GaSe under pressurization, **Figures 7(e)** and **7(f)** show the changing of $R(\omega)$ and $A(\omega)$ under pressure, respectively. These spectra remain insensitive to external pressure below the critical value P ~ 17 GPa. However, beyond this pressure, the curves change dramatically. Most incident photons are reflected back into the vacuum at the front surface, while a finite amount of light is absorbed even at low infrared frequency, attributable to free carrier concentration. The drastic decline of $R(\omega)$ and $A(\omega)$ indicates the plasmon modes discussed earlier. The behaviors have been verified by the previous reflectance measurements [36].

Hydrostatic pressure has emerged as an effective tool for tailoring the electronic and optical properties of layered materials for specific application demands, such as electronic, optoelectronic, and photovoltaic devices.

## 3.5 Phase transition mechanism

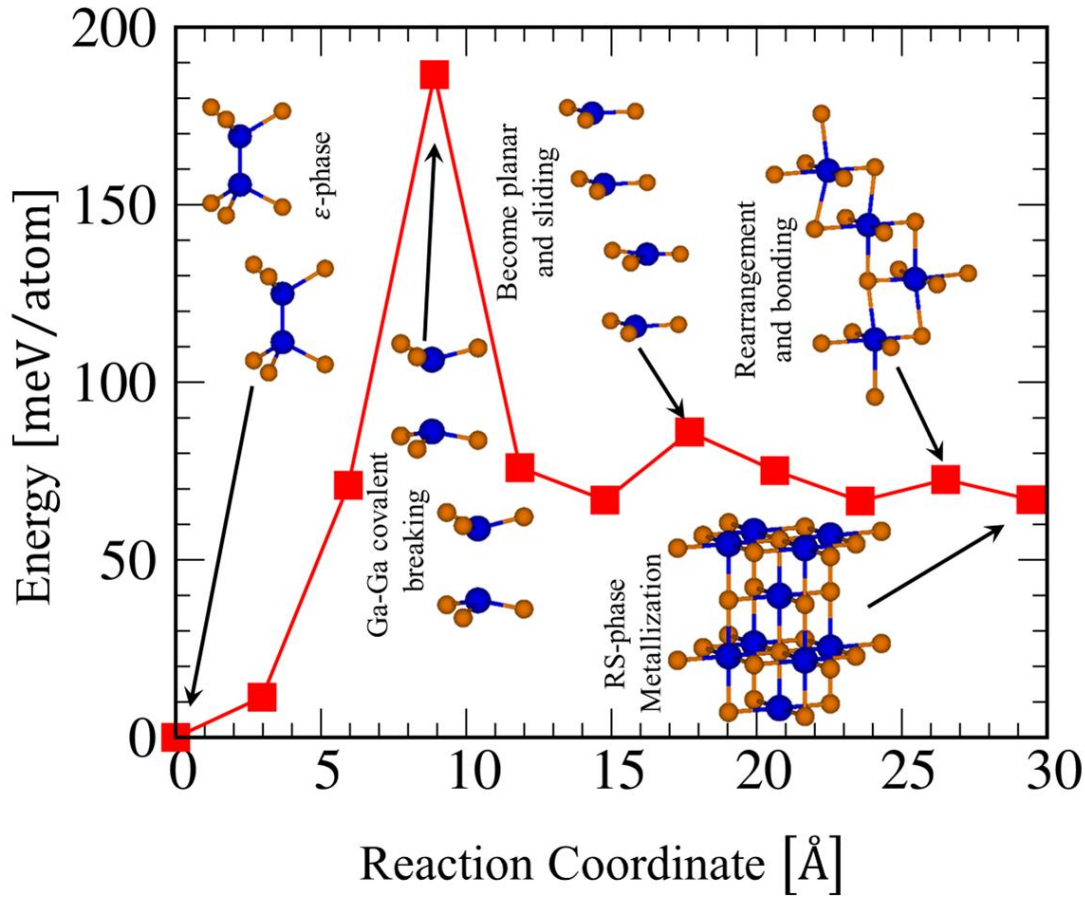

**Figure 8.** The proposed enthalpy barrier for GaSe Hexagonal-to-rocksalt transformation at the equilibrium pressure P = 17 GPa.

The SS-NEB calculation method, which is a commonly used technique to investigate transition states and reaction pathways in materials science, was utilized to extensively investigate the mechanism behind the structural phase transformation of GaSe. The calculation employed 10 intermediate images, with the root-mean-square force convergence set at 0.03 eV/Å. The process of the hexagonal-to-rocksalt phase transition can be regarded as a two-stage process in **Figure 8**.

During the first stage, the axial ratio c/a gradually decreases and the Ga-Ga covalent bonding is broken due to its instability under compressive forces. Ultimately, the bulk structure of GaSe becomes planar (intermediate state) resembling that of h-BN. In the second stage, the hexagonal angle $\gamma$ opens from $60^o$ to $90^o$ and two adjacent half-rigid layers slide alternately along the basal plane, and eventually the Ga and Se atoms from these layers come into contact and form metallic bonds, leading to the formation of an octahedral in the rocksalt GaSe structure through rearrangement and bonding. It is noteworthy that this transformation is accompanied by a shift from a semiconductor to a metallic transition, which is referred to as RS-phase metallization. The minimum energy required for the hexagonal-to-rocksalt transition has been confirmed to be approximately 200 meV per formula unit, highlighting the significant amount of energy necessary to break the strong Ga-Ga covalent bonds.

# 4. Conclusions

In conclusion, this study provides a comprehensive investigation of the phase transition, atomic vibration, electronic and optical properties, and excitonic effects of ε-GaSe under external pressure up to 40 GPa. The first-principles calculations show that ε-GaSe undergoes a structural phase transition from the hexagonal (ε) phase to the rock-salt (RS) phase at around 17 GPa, consistent with previous experimental and theoretical studies. The study reveals that the electronic properties of ε-GaSe undergo a semiconductor-metal transition at the transition point, with a sharp drop in the band gap due to the shift of the valence band maximum at the Γ-point and the conduction band minimum at the L- and H-point in the Brillouin zone.

Additionally, the study analyzed the phonon vibrations of ε-GaSe and found that the phonon frequencies increase gradually with increasing pressure up to the transition point, mostly due to the increased interlayer interactions and reduced intralayer bondings. The optical properties of ε-GaSe were also analyzed, revealing that the optical gap increases gradually with increasing pressure up to the transition point, and then drops sharply at the transition, in agreement with the electronic band gap. Moreover, the excitonic effects, such as the exciton binding energy and the oscillator strength, were found to be highly sensitive to pressure, with a significant drop to 0 eV observed at the transition pressure.

Finally, the solid-state Nudged Elastic Band (SS-NEB) method was used to analyze the phase transition mechanism of ε-GaSe, which showed that the transition involves a two-step process with an intermediate state before the final transformation to the RS phase. The activation energy for the transformation was estimated to be around 200 meV. The results obtained from this study provide new insights into the phase transition mechanism of ε-GaSe and can be useful in designing new materials for electronic and optoelectronic applications. Further experimental studies could validate these findings and pave the way for practical applications.

# Competing interests

The authors declare no competing interests.